\documentstyle[prd,aps,floats,epsfig]{revtex} 
 
\begin{document}  
\author{Carlo Contaldi$^{1}$, Mark Hindmarsh$^2$, and  
Jo\~ao Magueijo$^1$} 
\address{$^1$Theoretical Physics, The Blackett Laboratory, 
Imperial College, Prince Consort Road, London, SW7 2BZ, U.K.\\ 
$^2$ Centre for Theoretical Physics, University of Sussex,  
Brighton BN1 9QJ, U.K.} 
\twocolumn[\hsize\textwidth\columnwidth\hsize\csname@twocolumnfalse\endcsname 
 
\title{CMB and density fluctuations 
from strings plus inflation} 
 
\maketitle 
\begin{abstract} 
In cosmological models where local cosmic strings are  
formed at the end of a period of inflation, the perturbations are 
seeded both by the defects and by the quantum fluctuations. In 
a subset of these models, for example those based on $D$-term inflation,
the amplitudes are similar. Using our recent calculations of 
structure formation with cosmic strings,
we point out that in a flat cosmology with zero cosmological constant  
and 5\% baryonic component, strings plus inflation fits 
the observational data much better than each component individually.  
The large-angle CMB spectrum is mildly tilted, for Harrison-Zeldovich
inflationary fluctuations. It then  rises to a thick 
Doppler bump, covering $\ell=200-600$, modulated by soft secondary 
undulations.  The standard CDM anti-biasing problem is cured, 
giving place to a slightly biased scenario of galaxy formation. 
\end{abstract} 
\date{\today} 
] 
 
The combination of recent data from cosmic microwave background  (CMB)
observations and large scale structure (LSS) surveys is proving a  
strong discriminant amongst cosmologies.  The  
standard cold dark matter (sCDM) inflationary model \cite{scdm},  
dominant for so long, is now disfavoured as it predicts too high  
an amplitude for the CDM power spectrum at scales below about  
30 $h^{-1}$ Mpc \cite{infl}.  Cosmic strings and other  
topological defects \cite{csreviews},  
after resisting theoretical attack,
have also failed the tests  
in the standard cosmology of $\Omega=1$, $\Omega_\Lambda=0$,   
$\Omega_b = 0.05$ and $H_0 = 50$ km s$^{-1}$ Mpc$^{-1}$, as they  
cannot produce the required power on large scales \cite{pst,abr,all+,chm1}. 
Other cosmological parameters have been investigated in both  
inflationary \cite{infl} and defect \cite{otherdef} models, with  
the result that a cosmological constant improves matters in both  
scenarios.  The recent Type 1a Supernova data \cite{sn1a} supports this  
move away from an Einstein-de Sitter Universe.   
 
However, the failings of the inflationary and defect sCDM models are  
to a certain extent complementary, and an obvious question is whether 
they can help each other to improve the fit to CMB and LSS data.  Using 
our recent calculations for local cosmic strings \cite{chm1} and the by 
now familiar inflationary calculations \cite{cmbfast}, we are able to 
demonstrate that the answer is yes. Even with Harrison-Zeldovich
initial conditions and no inflation produced gravitational waves,
the large-angle CMB spectrum is mildly tilted, as preferred by COBE
data \cite{kris}. The CMB spectrum then rises into a thick Doppler bump, 
covering the region $\ell=200-600$, modulated by soft secondary 
undulations. More importantly the standard CDM anti-biasing problem is cured, 
giving place to a slightly biased scenario of galaxy formation.

 
It may seem baroque to invoke both inflation and strings to explain the 
cosmological perturbations. However, there is a very attractive inflationary  
model, namely $D$-term inflation \cite{Cas+89,DTermInfl},  
which necessarily produces strings, and in which  
the perturbation amplitudes are of similar amplitude \cite{rachel}
(see also \cite{LytRio98} for a review).   
$D$-term inflation requires the existence of an extra gauged U(1) symmetry,  
which is broken at the end of inflation, thereby resulting in the formation  
of a network of cosmic strings.   If the symmetry is not anomalous,
the strings will be local, otherwise they are global \cite{Cas+89}.
The inflaton corresponds to a flat direction in the potential,  
where the  energy density is set by the U(1) symmetry-breaking scale.   
Radiative corrections lift the flatness and force the fields eventually  
to the U(1)-breaking supersymmetric minimum.  A big attraction of 
the model is its naturalness in the technical sense: 
the flat direction is present as a result of  
symmetry, and the model avoids having to fine-tune any coupling constants.  

The strings appear as a result of the breaking of the U(1) symmetry, 
although the details of the process are far from certain.  It may be 
that the evolution of the homogeneous inflaton causes the U(1)-symmetric 
point in the potential to become unstable \cite{Cop+94}, or 
there may also be a non-thermal phase transition after inflation ends, 
during a ``preheating'' phase \cite{Tka+98}.  However, subject to the 
condition that the fields making the strings are uncorrelated at large 
distances, the subsequent evolution of the string network is thought to be 
independent of the formation process \cite{csreviews}.  
 
The ensuing structure formation scenario is highly exotic, and
worth studing just by itself. Regarded in abstract, 
structure formation may be due to two types of mechanism: 
active and passive perturbations. Passive fluctuations are due 
to an apparently acausal imprint in the initial conditions of the standard 
cosmic ingredients, which are then left to evolve by themselves. 
Active perturbations are due to an extra cosmic component, which 
evolves causally (and often non-linearly), and drives perturbations 
in the standard cosmic ingredients at all times.  
Inflationary fluctuations are passive. Defects are the  
quintessential active fluctuation. 
 
A scenario combining active and passive perturbation would 
bypass  most of the current wisdom on what to expect in either 
scenario. It is believed that the presence or absence of secondary 
Doppler peaks in the CMB power spectrum tests the very fundamental nature
of inflation, whatever its guise \cite{andbar}.
In the mixed scenarios we shall consider inflationary scenarios
could produce spectra with any degree of secondary oscillation
softening.

We shall now recap some of the main features of the 
$D$-term inflation model in which the strings plus inflation
scenario finds an attractive expression. To begin with, we 
define the reduced Planck mass $M = 1/\sqrt{8\pi G}$.  We 
recall that a supergravity theory is defined by two functions of the  
chiral superfields $\Phi_i$: the function $G(\bar\Phi,\Phi)$, 
which is related to the K\"ahler potential $K(\bar\Phi,\Phi)$ 
and the superpotential $W(\Phi)$ 
by $G=K+M^2\ln|W|^2/M^6$, and the 
gauge kinetic function $f_{AB}(\bar\Phi,\Phi)$.  
The scalar potential $V$ is composed of two terms, 
the $F$-term 
\begin{equation} 
V_F = M^2 e^{G/M^2}\left(G_i (G^{-1})^i_{\,j} G^j - 3 {M^2} \right)  
\end{equation}
and the $D$-term
\begin{equation}
V_D = \frac{1}{2} g^2 {\mathrm{Re}} f_{AB}^{-1} D^A D^B 
\end{equation} 
where  $g$ is the U(1) gauge  
coupling, $G^i = \partial G/\partial \Phi_i$, and 
$G_i = \partial G/\partial \bar\Phi^i$.  The function   
$D^A$ i given by 
\begin{eqnarray} 
D^A &=& G^i(T^A)_i^{\,j}\phi_j + \xi^A, 
\end{eqnarray} 
where the Fayet-Iliopoulos terms $\xi^A$, which we take to be 
positive, can be non-zero only for those $(T^A)_i^{\,j}$ which are U(1) 
generators.  We see that in order to have a positive potential energy density,  
either the $F$ term or the $D$ term must be non-zero.  In order to have
inflation, there must be a region in field space where the slow-roll
conditions $\epsilon \equiv \frac12 M^2 |V^i/V|^2 \ll 1$ and 
$|\eta| \equiv |\min {\mathrm eig} \, M^2 V^i_{\,j}/V| \ll 1$ 
are satisfied, where by the notation in the 
second condition we mean that the smallest 
eigenvalue of the matrix is much less than unity.
In $D$-term inflation, the conditions are satisfied because 
the fields move along a trajectory for which  
$\exp(G/M^2)$, $G^i$ and $G^i(T^A)_i^{\,j}\phi_j$ all vanish, leaving a tree-level  
potential energy density of $g^2\xi^A\xi^A/2$.  Thus the potential is  
completely flat before radiative corrections are taken into account.  At the  
end of inflation, if the fields are to relax to the supersymmetric minimum with  
$D^A+\xi^A = 0$, the U(1) gauge symmetries are necessarily broken,
assuming their  
corresponding Fayet-Iliopoulos terms are non-zero.  Thus strings are  
inevitable: the only question is how much inflation there is before the  
fields attain the minimum.

The simplest model \cite{DTermInfl} has three chiral fields,  
$S$, $\Phi_+$ and $\Phi_-$, which have charges 0, 1 and -1 under  
an extra U(1)$_X$ symmetry.  If one imposes an $R$-symmetry, the only  
possible superpotential is 
$W = \lambda S\Phi_+\Phi_-,$
and one assumes a K\"ahler potential with minimal form 
$K = |S|^2 + |\Phi_+|^2+|\Phi_-|^2$.
The scalar potential  
for the bosonic components $s,\phi_+,\phi_-$ is then 
\begin{eqnarray} 
V &=& \lambda^2|s|^2(|\phi_+|^2 + |\phi_-|^2) + \lambda^2 |\phi_+\phi_-|^2  
\nonumber\\
&+& \frac12 g^2(|\phi_+|^2 - |\phi_-|^2 + \xi)^2. 
\end{eqnarray} 
The vacuum $s=\phi_+=0$, $|\phi_-| = \sqrt\xi$ is supersymmetric but  
breaks the U(1) symmetry.  The field $s$ is massless at tree level, while 
the fields $\phi_\pm$ have masses $m^2_\pm = \lambda^2|s|^2 \pm g^2\xi$.  
Thus for $|s| > g\sqrt{\xi}/\lambda$, and $\phi_\pm =0$, the potential  
is flat in the $s$ direction and has positive curvature in the $\phi_\pm$  
directions. As a consequence of the broken supersymmetry, there  
are radiative corrections to the  potential along this  
flat direction from $\phi_\pm$ and their  
fermionic partners, resulting in an effective potential \cite{DTermInfl} 
\begin{equation} 
V_{\mathrm{eff}} = \frac12 g^2 \xi^2 \left(1 + \frac{g^2}{16\pi^2}  
\ln \frac{\lambda^2|s|^2}{\Lambda^2}\right), 
\end{equation} 
where $\Lambda$ is the renormalisation scale.  

It is also possible that the U(1) symmetry is anomalous.  However, 
the gauged U(1) symmetry is already effectively broken, leaving 
behind a global U(1) symmetry, which is broken when the charged fields 
gain expectation values.  This results in the formation of global strings 
\cite{Cas+89},
to which our calculations do not strictly apply.
 
The two major pieces of data to which we wish to compare the theory are  
the mean square temperature fluctuation in the multipole $\ell$, or 
$\ell(\ell+1)C_\ell/2\pi$, and the power spectrum of matter density  
fluctuations $P(k)={\langle |\delta(k)|\rangle}^2$.   
In order to compute CMB and LSS power spectra we note that fluctuations 
due to cosmic strings are imprinted long after their formation. The string 
network is produced at the end of inflation, and it is conceivable that 
the inflationary fluctuations and the initial configuration of strings 
are correlated. However strings are highly incoherent\cite{inc}, meaning 
that all string modes become decorrelated with themselves in time. 
Incoherence is due to the non-linear interactions present in the string 
evolutions, which lead to any Fourier mode being driven by all others. 
Hence the string network which produces the CMB and LSS fluctuations is 
surely uncorrelated with the string network produced at the end
of inflation and therefore with the inflationary fluctuations.  
 
The evolution of radiation, neutrinos, CDM, and baryons is 
linear for both string driven and inflationary perturbations. 
The fact that these two types of fluctuations are uncorrelated means 
that we can simply add the power spectra in CMB and LSS produced by 
each component separately. 
 
The spectrum of the perturbations from $D$-term inflation is  
calculable \cite{rachel}, and can be expressed in terms of $N$, 
the number of $e$-foldings between the horizon exit of cosmological  
scales today and the end of inflation, which occurs at $|\eta| = 1$. 
One finds  
\begin{equation} 
{\ell(\ell+1)C_\ell^{\mathrm{I}}\over 2\pi T_{\mathrm{CMB}}^2 }\simeq 
\frac14 |\delta_H(k)|^2 \simeq  
\frac{(2N+1)}{75} \left(\frac{\xi^2}{M^4}\right), 
\end{equation} 
where $ T_{\mathrm{CMB}}= 2.728 K$ 
is the temperature of the microwave  
background, and $\delta_H(k)$ is the matter perturbation amplitude at 
horizon crossing. The corrections to this formula, which is zeroth order in  
slow roll parameters, are not more than a few per cent.   
The inflationary fluctuations in this model are almost 
scale-invariant (Harrison-Zeldovich) and have a negligible tensor
component \cite{DTermInfl}. 

The string contribution is uncorrelated with the inflationary  
one, and is proportional to $(G\mu)^2$, where $\mu$ is 
the string mass per unit length, given by $\mu = 2\pi\xi$.  We can write  
it as  
\begin{equation} 
{\ell(\ell+1)C_\ell^{\mathrm{S}}\over 2\pi T_{\mathrm{CMB}}^2} =  
 \frac{{\cal A}^{\mathrm{S}}(\ell)}{16}\left(\frac{\xi^2}{M^4}\right), 
\end{equation} 
where the function ${\cal A}^{\mathrm{S}}(\ell)$ gives the amplitude of the  
fractional temperature fluctuations in units of $(G\mu)^2$.  Allen et al.\  
\cite{all+} report ${\cal A}^{\mathrm{S}}(\ell) \simeq 60$ on large  
angular scales, with little dependence on $\ell$.  Our simulations give  
${\cal A}^{\mathrm{S}}(\ell) \simeq 120$, with a fairly strong tilt.
The source of the difference is not alogether clear: our simulations 
are based on a flat  
space code which neglects the energy losses of the strings through  
Hubble damping.  The simulations of Allen et al.\ do include Hubble damping, 
which would tend to reduce the string density and hence the normalisation. 
However, they have a problem of lack of dynamic range, 
and therefore may be missing  
some power from strings at early times, and therefore higher $\ell$.
 
Jeannerot \cite{rachel} took the Allen--Shellard normalisation 
and $N\simeq 60$, and found that the  
proportion of strings to inflation is roughly $3:1$.  
With our normalisation, the approximate ratio is $4:1$.
In any case this ratio is far  from a robust prediction in strings 
plus inflation
models, as it depends on the number of $e$-foldings,
and the string normalisation, both of which are uncertain.  
We will therefore leave it as a free parameter. For definiteness
we shall parametrize the contribution due to strings and inflation
by the strings to inflation ratio $R_{\mathrm{SI}}$, 
defined as the ratio in $C_\ell$ at $\ell=5$, that is
$R_{\mathrm{SI}}=C_5^S/C_5^I$. 
 
In Figs. 1 and 2 we present power spectra in CMB and CDM produced 
by a sCDM scenario, by cosmic strings, and by strings plus inflation. 
We have assumed the traditional choice of parameters, setting the Hubble 
parameter
$H_0=50$ km sec$^{-1}$ Mpc$^{-1}$, the baryon fraction to
$\Omega_b=0.05$, and  assumed a flat geometry, no cosmological
constant, 3 massless neutrinos, standard recombination,
and cold dark matter. The inflationary perturbations have a 
Harrison-Zeldovich or scale invariant spectrum, and the amount of
gravitational radiation (tensor modes)
produced during inflation is assumed to be negligible.
The cosmic strings are assumed to attain scaling by losing their energy
into gravitational radiation, or some other non-interacting radiation fluid. 
Other assumptions for the equation of state of the decay products may be 
made \cite{chm1}: however, a relativistic equation of state produces the
worst bias problem at large scales and thus represents the ``worst case''
string scenario.

\begin{figure} 
\centerline{\psfig{file=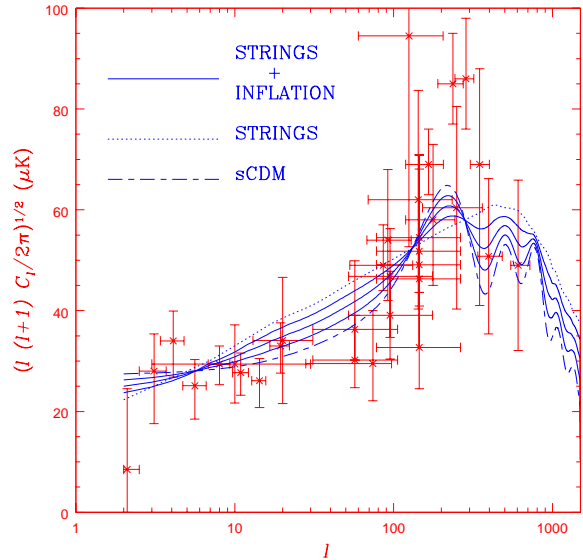,width=8 cm,angle=0}} 
\caption{The CMB power spectra predicted by cosmic strings, sCDM, 
and by inflation and strings with
$ R_{\mathrm{SI}}=0.25,0.5,0.75.$ 
The large angle
spectrum is always slightly tilted. The Doppler peak becomes a thick
Doppler bump at $\ell=200-600$, modulated by mild undulations.} 
\label{fig1} 
\end{figure}

We now highlight the main features in the resulting CMB and LSS power
spectra. The CMB power spectrum shape in these models
is highly exotic. The inflationary 
contribution is close to being Harrison-Zeldovich. Hence it produces
a flat small $\ell$ CMB spectrum. The admixture of strings, 
however, imparts a tilt.
Depending on $R_{\mathrm{SI}}$ one may tune the CMB plateau tilt between 1
and about 1.4, without invoking primordial tilt and inflation
produced gravity waves.
This may be seen as a positive feature, as a flat spectrum is 
not favoured by the COBE data \cite{kris}. 

The proverbial inflationary Doppler peaks are transfigured in these
scenarios into a thick Doppler bump, covering the region  
$\ell=200-600$. The height of the peak is similar for sCDM and strings,
with standard cosmological parameters. Fitting two of the Saskatoon
points, therefore  requires fiddling with cosmological parameters.
The Doppler bump is modulated by small undulations, 
which cannot truly be called secondary peaks. 
By tuning $R_{\mathrm{SI}}$ one may achieve any degree
of secondary oscillation softening. This provides a major
loophole in the argument linking inflation with secondary
oscillations in the CMB power spectrum. If these oscillations were not
observed, inflation could still survive, in the form of the models
discussed in this Letter.

The CMB fluctuations in these models combine a Gaussian component,
produced by inflationary fluctuations, and a non-Gaussian component,
due to strings. The superposition of Gaussian and  
non-Gaussian maps often leads to rather subtle non-Gaussian
structures, visually indistinguishable 
from Gaussian maps \cite{fermag}. 
Sophisticated statistics would certainly be required to 
recognize the non-Gaussian signal in these theories\cite{fermag,alex}.

In these scenarios the LSS of 
the Universe is almost all produced by inflationary fluctuations. 
However COBE scale CMB anisotropies are due to both strings and  
inflation. Therefore COBE normalized CDM fluctuations are  
reduced by a factor $(1+R_{\mathrm{SI}})$ in strings plus inflation scenarios. 
This is equivalent to multiplying 
the sCDM bias by ${\sqrt{1+R_{\mathrm{SI}}}}$ on all scales, except
the smallest, where the string contribution may be
non negligible. Given that 
sCDM scenarios produce too much structure on small scales
(too many clusters)  
this is a desirable feature.

We find that the the bias required to fit the power spectrum 
of Peacock and Dodds \cite{pdodds} at the 8 $h^{-1}$Mpc scale is
$b_{8}=\sigma_{8}^{\mathrm{PD}}/\sigma_{8}=0.7,0.8,1.0$, 
for $R_{\mathrm{SI}}=0.25,0.5,0.75$ respectively.
In 100 $h^{-1}$Mpc spheres one requires bias $b_{100}=
\sigma_{100}^{\mathrm{PD}}/\sigma_{100}=1.0, 1.2, 1.6$
to match observations. None of these values is uncomfortable.
If most of the objects used to estimate $P(k)$ evolved from
high peaks of the density field, one should have biasing,
not anti-biasing, and the bias should also be of order 1.
Strings plus inflation with $R_{\mathrm{SI}}>0.75$
complies with these requirements,
whereas each separate component does not.
However, 
with the simplest choice of cosmological parameters,
the bias must be  scale dependent in these models.

The model naturally inherits most 
of the sCDM good features. For instance, it fits the 
Lyman break galaxy clustering found in \cite{steidel}, and 
the constraints inferred from damped Lyman-$\alpha$ systems 
\cite{lalfa}.

\begin{figure} 
\centerline{\psfig{file=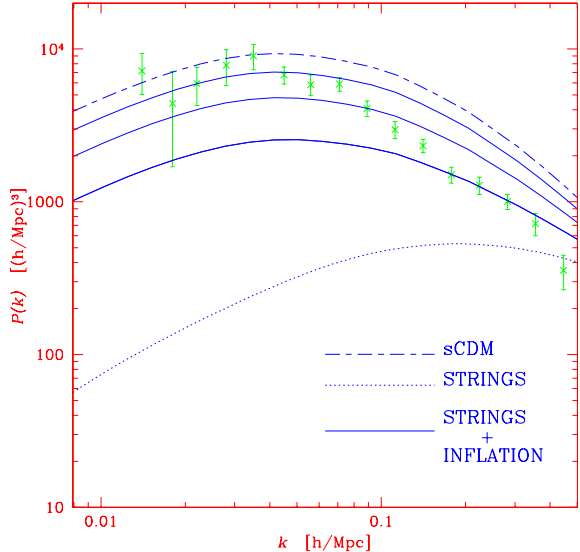,width=8 cm,angle=0}} 
\caption{The  power spectra in CDM fluctuations 
predicted by cosmic strings, sCDM, 
and by inflation and strings with $R_{\mathrm{SI}}=0.25,0.5,0.75.$} 
\label{fig2} 
\end{figure}

If we are prepared to combine inflation with 
more unusual string scenarios,
in which, say, the strings loose 
energy in a direct channel 
into CDM \cite{chm1}, then the major novelty is that
strings will be responsible for the LSS on scales below
about 30$h^{-1}$Mpc. We will discuss elsewhere how this may
have interesting implications for the time evolution of the CDM
power spectrum \cite{steidel}. Active models drive fluctuations
at all times, and therefore produce a time-dependence
in $P(k)$ different from passive models. In such models there
would also be intrinsic non-Gaussianity at the scale of clusters,
with interesting connections with the work of \cite{james}.

In summary, mixing cosmic string and sCDM spectra smoothes 
the hard edges of either separate component, leaving 
a much better fit to LSS and CMB power spectra (see \cite{sug}
for another happy marriage). 
It is perhaps rather ironic that inflation and strings, often  
presented in opposition to one another, should find such a fruitful 
union.  We do not wish to claim that the outcome is perfect: the shape 
of the power spectrum still goes wrong in the 
standard cosmology.  The purpose of this work is to start the 
investigation of a new set of cosmological models, those which combine 
inflation and defects. As the data improves we will be able to constrain 
them more heavily, particularly as the first peak in the CMB power 
spectrum begins to be traced.  Meanwhile, we are left with an intriguing 
hybrid model.

We thank Rachel Jeannerot for urging 
us to carry out this project, and R.Battye, P. Ferreira, A.
Liddle and J. Weller for helpful comments.
We acknowledge financial support from the Beit Foundation (CC),  
PPARC (MH), and the Royal Society (JM).

\end{document}